\documentclass[acmsmall,screen]{acmart}

\AtBeginDocument{%
  \providecommand\BibTeX{{%
    \normalfont B\kern-0.5em{\scshape i\kern-0.25em b}\kern-0.8em\TeX}}}

\setcopyright{acmcopyright}
\copyrightyear{2018}
\acmYear{2018}
\acmDOI{XXXXXXX.XXXXXXX}

\usepackage{soul}
\usepackage{multirow}
\usepackage{listings}
\usepackage{fancybox}
\usepackage{enumitem}
\usepackage{subcaption}
\usepackage{color}
\usepackage{xcolor}
\usepackage{array}
\usepackage{amsmath}
\usepackage{graphicx}
\usepackage{xspace}
\usepackage{url}
\usepackage{tikz}
\usepackage{caption}
\usepackage{pgfplots}
\pgfplotsset{compat=1.16}
\usepackage{pgf-pie}
\usetikzlibrary{pgfplots.statistics,calc}
\usepackage{pifont}
\usepackage{tabularx}
\usepackage[ruled,vlined]{algorithm2e}
\usepackage{graphicx}
\usepackage{array}
\usepackage{amsmath}
\usepackage{hyperref}
\usepackage{multirow}
\usepackage{caption}
\usepackage{flushend}
\usepackage{balance}
\usepackage{ulem}
\usepackage{color, colortbl}
 \usepackage{enumitem}
 \usepackage{balance}
\usepackage{booktabs}
\usepackage{fancyhdr}
\usepackage[framemethod=TikZ]{mdframed}
\usepackage{tcolorbox}
\makeatletter
\newcommand{\mybox}[4]{
    \begin{figure}[H]
        \centering
    \begin{tikzpicture}
        \node[anchor=text,text width=\columnwidth-1.2cm, draw, rounded corners, line width=1pt, fill=#3, inner sep=5mm] (big) {\\#4};
        \node[draw, rounded corners, line width=.5pt, fill=#2, anchor=west, xshift=5mm] (small) at (big.north west) {#1};
    \end{tikzpicture}
    \end{figure}
}

\usepackage{tcolorbox}

\definecolor{songcolor}{RGB}{191,191,191}

\definecolor{darkpastelgreen}{rgb}{0.01, 0.75, 0.24}
\definecolor{darkseagreen}{rgb}{0.56, 0.74, 0.56}

\newcommand{\RACK}{\texttt{RACK}}

\newcommand{\cmark}{\ding{51}}%
\newcommand{\xmark}{\ding{55}}%

\begin{document}

\title{A Survey on Query-based API Recommendation}

\author{Moshi Wei}
 \email{moshiwei@yorku.ca}
 \orcid{0000-0003-1659-1960}
 \affiliation{%
   \institution{York University}
   \streetaddress{4700 Keele St.}
   \city{North York}
   \state{Ontario}
   \country{Canada}
   \postcode{M3J 1P3}
 }

\author{Nima Shiri harzevili}
\email{nshiri@yorku.ca}
\affiliation{%
  \institution{York University}
  \streetaddress{4700 Keele St.}
  \city{North York}
  \state{Ontario}
  \country{Canada}
  \postcode{M3J 1P3}
}

\author{Alvine Boaye Belle}
\affiliation{%
  \institution{York University}
  \streetaddress{4700 Keele St.}
  \city{North York}
  \country{Canada}}
\email{alvine.belle@lassonde.yorku.ca}

\author{Junjie Wang}
\email{junjie@iscas.ac.cn}
\orcid{0000-0002-9941-6713}
\affiliation{%
  \institution{Institute of Software Chinese Academy of Sciences}
  \streetaddress{4\# South Fourth Street, Zhong Guan Cun}
  \city{Beijing}
  \country{China}
  \postcode{100190}
}

\author{Lin Shi}
\email{shilin@buaa.edu.cn}
\orcid{0000-0003-1476-7213}
\affiliation{%
  \institution{ Institute of Software Chinese Academy of Sciences}
  \streetaddress{4\# South Fourth Street, Zhong Guan Cun}
  \city{Beijing}
  \country{China}
  \postcode{100190}
}

\author{Jinqiu Yang}
\affiliation{%
  \institution{Concordia University}
  \streetaddress{1455 Blvd. De Maisonneuve Ouest, Montreal, Quebec H3G 1M8}
  \city{Montreal}
  \country{Canada}}
\email{jinqiu.yang@concordia.ca}

\author{Song Wang}
\affiliation{%
  \institution{York University}
  \streetaddress{4700 Keele St.}
  \city{North York}
  \country{Canada}}
\email{wangsong@yorku.ca}

\author{Zhen Ming (Jack) Jiang}
\affiliation{%
  \institution{York University}
  \streetaddress{4700 Keele St.}
  \city{North York}
  \country{Canada}}
  \email{zmjiang@eecs.yorku.ca}

\begin{abstract}
Application Programming Interfaces (APIs) are designed to help developers build software more effectively. Recommending the right APIs for specific tasks has gained increasing attention among researchers and developers in recent years.  
To comprehensively understand this research domain, we have conducted a survey to analyze API recommendation studies published in the last 10 years. 
Our study begins with an overview of the structure of API recommendation tools. Subsequently, we systematically analyze prior research and pose four key research questions. For RQ1, we examine the volume of published papers and the venues in which these papers appear within the API recommendation field. In RQ2, we categorize and summarize the prevalent data sources and collection methods employed in API recommendation research. In RQ3, we explore the types of data and common data representations utilized by API recommendation approaches. We also investigate the typical data extraction procedures and collection approaches employed by the existing approaches. 
RQ4 delves into the modeling techniques employed by API recommendation approaches, encompassing both statistical and deep learning models.  Additionally, we compile an overview of the prevalent ranking strategies and evaluation metrics used for assessing API recommendation tools. Drawing from our survey findings, we identify current challenges in API recommendation research that warrant further exploration, along with potential avenues for future research. 

\end{abstract}



\maketitle

\section{Introduction}
\label{Introduction2}

Choosing the appropriate software Application Programming Interfaces (API) is crucial in ensuring the successful implementation of software projects. API recommendation tools provide a service that identifies and suggests the appropriate APIs to use, based on the specific needs and objectives of a project. This systematic literature review investigates the latest developments, progress, and empirical results in API method recommendations for software engineering developers. The aim is to recommend APIs to developers, and the survey highlights gaps in the research literature in this field. Based on the identified gaps and technological opportunities in the literature, the article also identifies open issues and challenges faced by API method recommendations.

The field of API recommendations has undergone significant development and refinement over time~\cite{gu2016deep,rahman2016rack,wei2022clear,allamanis2018survey,feng2020codebert}. New computer technologies and algorithms have had a profound impact on API recommendations. The underlying technologies for API recommendations have evolved from early API usage pattern analysis~\cite{allamanis2013mining,chan2012searching} to later deep learning-based API recommendations~\cite{feng2020codebert,gu2016deep} and most recently, API recommendations based on large language models~\cite{kang2021apirecx}. Despite significant changes in the API recommendations field, certain patterns have been identified that are helpful for future research directions and address pending technical challenges, especially through recognizing patterns from existing studies. For example, many studies find that relying solely on the information provided in API documentation is insufficient for accurate and comprehensive API recommendations~\cite{chen2021holistic,gao2021api,gu2019codekernel,hussain2020codegru}. Consequently, considerable effort has been invested in collecting new data sources, with some typical sources being StackOverflow, GitHub, and the Android app store. 

A typical API recommendation approach often contains the following steps. 
\textbf{Data Collection:} The first step toward building an API recommendation model is to collect relevant query-API pair data. Multiple sources provide vulnerability detection datasets, such as GitHub~\cite{liu2019autoencoder, yan2018learning, nguyen2016learning}, StackOverflow ~\cite{huang2018api, zhang2017recommending, liu2023recommending}, and online documents ~\cite{nguyen2016api, gao2021api}. Researchers often conduct new data collection in experiments rather than reusing existing standard datasets due to the dynamic nature of software APIs, which continuously evolve through version updates. \textbf{API Usage Extraction and Query Extraction:} Once the data is collected, it needs to be processed into query-API pairs for modeling purposes. API extraction techniques fall into two groups based on the data source: single-API usage information and multi-API usage information. The latter is also named API usage patterns, representing sequences of frequent API method calls extracted from software source code. This process requires appropriate extraction techniques such as document parsing ~\cite{thung2013automatic, ajam2021scout, zhou2021boosting, yin2021api, kim2013enriching, lv2015codehow, xie2020api, huang2018api, hussain2020codegru, yuan2019api}, graph analysis ~\cite{sun2021task, zhao2021icon2code, nguyen2012grapacc, nguyen2015graph, chen2021holistic, gao2021api, ling2021graph}, and abstract syntax tree (AST) parsers~\cite{jones2003abstract, cai2019abstract, yamaguchi2012generalized, koyuncu2020fixminer}. 
\textbf{Modeling:} Modeling is the most evolved process in API recommendation. Techniques range from statistic-based approaches such as indexing ~\cite{tran2013api, xie2020api, heinemann2012identifier, rahman2016rack, nguyen2012grapacc, nguyen2016api, raghothaman2016swim} to machine learning approaches such as Word2Vec~\cite{liu2019design, nguyen2017exploring, lee2021mining, van2017combining}, and the latest deep learning techniques such as LSTM~\cite{tian2018automatically, yan2018learning, gu2016deep, gu2018deep}.  \textbf{Evaluation:} Once the modeling is finished, the model's performance is evaluated using a separate test dataset. Various metrics, such as accuracy, precision, recall, and F1 score, are calculated to assess the model's effectiveness in detecting vulnerabilities. The model may also be validated against real-world API queries to measure its practical utility.

There are several surveys related to API recommendations or similar directions. We collected a list of review papers~\cite{peng2022revisiting, xie2023survey, allamanis2018survey, shin2021survey, liu2021opportunities} related to API recommendations and carefully compared the differences of each work. The details of the comparison are described in Section~\ref{related}. 
We investigate the existing studies for API recommendation and proposed a list of research questions (RQs) to summarize our findings:

\begin{itemize}
\item \textbf{RQ1.} \textit{What are the trends in API recommendation studies?}

\begin{itemize}
  \item{What is the distribution of publication time in API recommendation studies?}
  \item{What is the distribution of publication venues in API recommendation studies?}
  \item{what are the frequent key directions and concepts in API recommendation studies?}
\end{itemize}
\item \textbf{RQ2.} \textit{What are the common data sources in API recommendation tools?}

\item \textbf{RQ3.} \textit{What are the common data processing techniques applied in API recommendation tools?} 

\begin{itemize}
  \item{What are the common API usage pattern types?}
  \item{What are the common API usage extraction methods?}
\end{itemize}

\item \textbf{RQ4.} \textit{What are the common modeling techniques in API recommendation tools?} 
\item \textbf{RQ5.} \textit{What are the common evaluation processes in API recommendation tools?} 
\end{itemize}

In this survey, 34 papers were collected from 19 conferences and journals between 2012 and 2023. Initially, a set of search terms was formulated using 10 API recommendation paper titles familiar to the research group. These terms guided exploration in various paper databases, and collected papers underwent manual and automatic filtering. Through comprehensive analysis, the survey addresses research questions (RQs) such as the main trends in API recommendation research, common components of existing API recommendation tools, unique features of current API recommendation tools, and typical evaluation processes in API recommendation research. This paper makes the following contributions:

\begin{itemize}

\item To the best of our knowledge, we are the first to systematically study natural language-based API recommendation systems and their subsystems, including data sources, data extraction, modeling, and evaluation metrics.

\item We thoroughly analyzed 34 relevant query-based API recommendation studies concerning publication trends and venues.

\item We provided a classification of modeling techniques used in API recommendation based on their architectures, along with an analysis of the technique selection strategy employed in these models.

\item We analyze the challenges inherent in the current state of API recommendation research and provide directions for future API recommendation research endeavors.

\item We have shared our results and analysis data as a replication package\footnote{10.5281/zenodo.10359962} to allow other researchers to follow this paper and extend it.

\end{itemize}

The structure of the paper is as follows: Section \ref{background} outlines typical API recommendation system processes by defining software engineering technical terms in the API recommendation research field and providing background information for the survey. Section \ref{related} introduces existing work related to the survey. Section \ref{Methodology} outlines the survey method, while Sections \ref{RQ1} to \ref{RQ5} provide detailed insights into each research question and its corresponding answers. In Section \ref{challenges}, challenges related to API recommendations are discussed, and Section \ref{Opportunities} explores potential opportunities. Section \ref{conclusion} concludes the study.

\section{Background}
\label{background}


\subsection{Application Programming Interface}

In the software engineering domain, an Application Programming Interface (API) represents a set of methods designed for communication between machines. Software libraries under active development usually update their API regularly, ensuring its alignment with evolving requirements and support of new features. The risk of software failure increases during this practice due to version mismatch, as the API in the latest version sometimes becomes incompatible with its codebase.

In 1991, Malamud~\cite{malamud1991analyzing} introduced the term API as ``\textit{a compilation of services accessible to programmers, intended for the execution of specific tasks.}'' The phrase ``\textit{API method call}'' refers to the act of invoking an API method within the source code. Although similar to conventional method invocations, API method calls have a unique attribute: APIs are external methods exposed to end-users for software development and integration. APIs consist of a set of methods, regulations, and protocols tailored to streamline communication among software applications. In contrast, a method refers to a general function within the source code responsible for executing a specific action or task, which can be both internal or external.

\subsection{API Usage Pattern Mining}
API usage pattern mining involves collecting an extensive repository of source code snippets and utilizing specialized tools to identify recurring usage patterns within these code fragments~\cite{zhong2009mapo}. The goal of API usage pattern mining is to identify recurring sequences of API method calls. For instance, when a developer wants to update a string in a file, an API usage pattern would provide recommendations that suggest the use of a file object to access the file, then create a file writer function for the string write operation, and finally call the close function of the file object after the operation to close the file. Within an API usage pattern, a user can expect at least one frequently occurring sequence of API method calls~\cite{wang2013mining}. Furthermore, API usage pattern mining techniques can be applied to real-time systems as well. Such systems require rapid response times and optimal memory efficiency to effectively fulfill their operational requirements. Therefore, it is critical to analyze the performance bottlenecks of their APIs through API usage pattern mining~\cite{linares2014mining}.

\subsection{Software Development Kit}
A Software Development Kit (SDK) represents a more comprehensive toolkit for crafting applications on specific platforms compared to an API, which primarily facilitates system-to-system communication. In addition to APIs, SDKs frequently encompass supplementary libraries, tools, documentation, and code samples. SDK significantly diminishes the complexity and learning curve associated with the seamless integration of third-party functionalities into applications. Notably, SDK developers usually maintain a community that provides code samples and tutorials to facilitate developers in gaining proficiency in utilizing the SDK's APIs~\cite{yin2021api}.

Furthermore, some SDKs offer functionalities that extend beyond the scope of basic functional and communicative APIs. For instance, the Android SDK, designed for Android application development, not only offers functional APIs but also delivers pre-defined UI components, debugging tools, and emulators to provide extensive support for application development, which demonstrates the depth of support SDKs can provide. SDKs commonly feature a comprehensive ecosystem where users can actively contribute by building and sharing customized modules. This community-driven aspect not only benefits the versatility of the SDK itself but also establishes a robust sense of community around its usage. This collaborative ecosystem becomes a valuable asset that nurtures innovation, sharing of insights, and collaborative problem-solving among developers.

\subsection{Abstract Syntax Tree}

The Abstract Syntax Tree (AST) functions as a tree representation of a code snippet's source code, effectively conveying the structural information of the source code. Each node within this tree representation corresponds to an entity present in the source code. Similar to the behavior of a source code compiler, the AST representation retains solely the structural and content-related attributes of the code while discarding all associated comments. Also, these ASTs are often used as input in expressing a program's source code structure for a static-based code analyzer. For example, in source code refactoring, developers can use the structural information in the ASTs to perform effortless transformations and refactor the code snippet. By examining the AST of a program without executing the code, automated code analysis tools can effectively detect syntax-related issues. In API recommendation systems, ASTs frequently serve as a foundational data representation, representing structural information from the source code~\cite{nguyen2016api,xi2020api,tian2018automatically,yan2018learning}.

\subsection{API Method Recommendation}

API method recommendation tools provide users with suggestions for appropriate API methods based on their input. Generally, there are two primary categories of API recommendation tools. The first is question-and-answer systems, which operate similar to intelligent search engines. These systems recommend specific API method names in response to programming queries submitted by users. For example, \textit{DeepAPI}~\cite{gu2016deep} proposed by Gu et. al. suggests an API method based on learning massive API sequence data using an LSTM model. When a user submits a programming query such as ``How can I sort an ArrayList in Java?'' to \textit{DeepAPI}, the tool calls its deep learning model to process the query and returns the suggested API method \texttt{ArrayList.sort} as the solution. This type of recommendation tool relies solely on the textual content input of the query to provide relevant suggestions.

The second category of API recommendation tools involves code completion utilities, designed to facilitate developers by offering recommendations for the next API method based on the context provided by the current API. An illustrative example is the code completion tool \textit{PyReco}~\cite{d2016collective}, which incorporates API recommendation capabilities. \textit{PyReco} operates in real-time, suggesting API methods that are contextually relevant to the code being written. For instance, when the source code contains statements like \texttt{file = "samples/sample\_file.txt"} and \texttt{f = open(file, "w")}, \textit{PyReco} intelligently suggests \texttt{f.readlines()} as the next API method to consider. Typically, these code completion tools seamlessly integrate into popular development environments such as IntelliJ IDEA or Visual Studio Code, enhancing the coding experience by offering timely and context-aware API recommendations.

\subsection{Overview of API Recommendation System}

\begin{figure*}[t!]
\centering
\includegraphics[width=\textwidth]{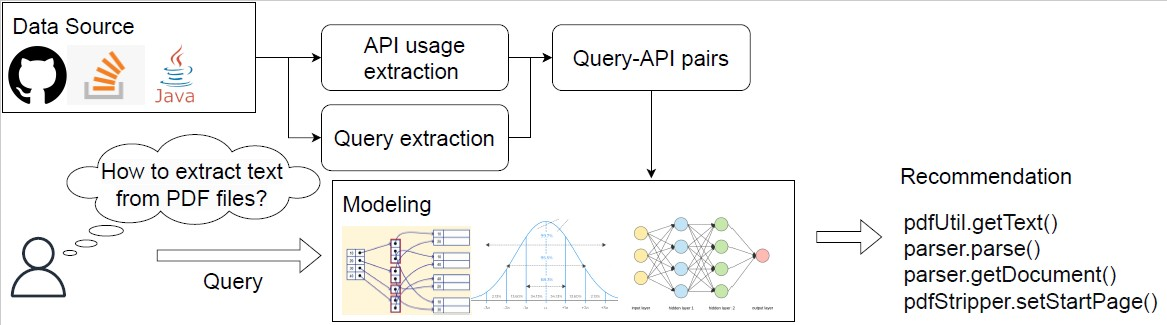}
\caption{The overview of query-based API recommendation}
\label{fig:overview}
\end{figure*}

Figure~\ref{fig:overview} illustrates the workflow of the API recommendation tool, comprising six key components: Data Source, API Usage Extraction, Query Extraction, Modeling, and Evaluation.

\textbf{Data Source}: Query-based API recommendation tools usually collect data from open source code repositories such as Github or online source code forums such as Stack Overflow. Such data sources are ideal for collecting query text and corresponding source code because they contain a substantial amount of API usage-related information~\cite{monperrus2016searching}. Researchers leverage multiple sources for building datasets for API recommendation, including Stack Overflow~\cite{rahman2016rack}, GitHub~\cite{gu2016deep,gu2018deep}, and Official API Documents~\cite{chen2021holistic},  to gather data for building the dataset. Most recommendation tool creates their specialized dataset tailored to the specific programming language or domain it serves. For instance, a Java API recommendation tool's knowledge base primarily contains information about Java API usage. Further details regarding the Data Source are discussed in Section \ref{RQ3_1}.

\textbf{API Usage Extraction and Query Extraction}: Before modeling, the API usage data collected in the Data Source requires processing. The choice of API usage extraction techniques depends on the modeling input ~\cite{maddison2014structured}. For instance, a graph-based API recommendation tool utilizes graph-based modeling and thus requires graph input. In such cases, the selected API usage extraction technique transforms the data in the knowledge base into a graph format. Conversely, many API recommendation tools adopt code processing techniques focused on extracting API method names from code snippets or documents~\cite{qi2020data, wang2018mashup, wang2013mining}. Common processing methods include Graph Analysis~\cite{nguyen2012grapacc,nguyen2015recommending,chen2021novel,liu2018effective}, and Abstract Syntax Tree (AST) parsing~\cite{ asaduzzaman2016simple,he2021pyart,yan2018learning,wang2021plot2api}. Please see the detailed discussion on API Usage Extraction in Section \ref{RQ3}.
The training phase of an API recommendation tool requires both query data and API data in pairs. The queries are collected from the same data source from which the code snippet is collected. Similar to API usage extraction, the query collected cannot be directly used in training, this is because users often have multiple ways of expression for the same question, and traditional tools may not support such flexibility in querying a recommendation system.



\textbf{Modeling}: Modeling serves as the core technique responsible for converting the instances or queries within a knowledge base into numerical representations for computation and comparison. Within the API recommendation tool, modeling represents a core component. Existing research in the field of API recommendation predominantly employs modeling techniques including statistics, Machine Learning (ML), and Deep learning (DL). The techniques for API recommendation models include TF-IDF vectors, Graph Neural Networks (GNNs)~\cite{chen2021holistic}, Collaborative Filtering (CF)~\cite{nguyen2021recommending}, Recurrent Neural Networks (RNNs)~\cite{fucci2019using}, and Transformers~\cite{wei2022clear}. A detailed discussion of Modeling can be found in Section \ref{RQ4}.



\textbf{Evaluation}: In general, API recommendation tools commonly employ evaluation metrics derived from information retrieval techniques. Since the majority of programming queries can be adequately addressed with just one or two APIs, most API recommendation tools prioritize retrieving the first or the initial few correct answers. Therefore, it is logical to evaluate tool performance using information retrieval metrics~\cite{nguyen2019focus}. For API recommendation tools that propose sequences of APIs, BLEU serves as a measure to assess the quality of the generated sequence relative to the desired answer~\cite{gu2016deep}. Additional details on the Evaluation aspect are provided in Section \ref{RQ5}.
\section{Related work}
\label{related}

\begin{table}[t!]
\caption{Comparison with existing surveys}
\label{table_20}
\centering
\resizebox{0.9\columnwidth}{!}{%
\begin{tabular}{lccccc}
\toprule
Paper                                     & Data source     & Evaluation & Model      & Representation & Direction \\
\midrule
Xie et al.\cite{xie2023survey}            & \checkmark      & \checkmark        & \checkmark & \checkmark     & Code Search         \\
Liu et al.\cite{liu2021opportunities}     &                 & \checkmark        & \checkmark &                & Code Search         \\
Allamanis et al.\cite{allamanis2018survey}&                 &                   & \checkmark & \checkmark     & Code Models         \\
Shin et al.\cite{shin2021survey}          &                 &                   &            & \checkmark     & Code Models         \\ 
Peng et al.\cite{peng2022revisiting}      & \checkmark      & \checkmark        &            &                & API Recommendation  \\
Ours                                      & \checkmark      & \checkmark        & \checkmark & \checkmark     & API Recommendation  \\

\bottomrule
\end{tabular}%
}
\end{table}

We carefully searched for existing API recommendation-related review papers which resulted in a collection of the following review papers. 
Peng et.al.~\cite{peng2022revisiting} surveyed Application Programming Interfaces (APIs) and found that the growing number of APIs has made it difficult for developers to find the appropriate APIs for their needs. To address this challenge, many researchers have focused on improving the API recommendation task, but the lack of a uniform definition and standardized benchmark has made it difficult to evaluate the performance of new models. This review aims to address this issue by benchmarking various approaches to API recommendation and identifying actionable insights and challenges for improving the performance of these models. Compared to our study, they focus on the benchmarking of existing tools, leaving other aspects such as the reasoning of data source collection and input and output forms of API recommendation unexplored. Compared to Peng et.al.'s work, our work focuses on the API recommendation at an extensive scale, exploring models and representations related to questions that are missing in previous work. 

Xie et.al.~\cite{xie2023survey} surveyed deep learning-based code search including several API recommendation-based approaches code search tools. such tools recommend APIs based on user input and then use API as the index for further code snippet retrieval. This survey focuses only on a deep learning-based approach and misses other important approaches such as graph-based approaches.

Allamanis et.al.~\cite{allamanis2018survey} conducted a survey of recent research at the intersection of machine learning, programming languages, and software engineering, which has proposed learnable probabilistic models of source code that exploit patterns in code. The survey contrasts programming languages with natural languages and discusses how their similarities and differences drive the design of probabilistic models. The authors present a taxonomy of the models based on their design principles and use it to review the literature, as well as discuss challenges and opportunities in applying these models to various application areas. This review explores the adaptation of these models to different applications, addressing challenges and opportunities in the process while our research focuses on API recommendation methods and applications.

Shin et.al.~\cite{shin2021survey} surveyed source code generation models for natural language inputs, and analyzed the current trend, suggesting future research directions, including customizing language models and exploring better source code representations. This review also discusses embedding techniques and pre-trained models for Code models. Such techniques have been successful in natural language processing tasks. Compared to our research, this review focuses more on the code generation task while our research focuses on API method recommendation.

Liu et al.~\cite{liu2021opportunities} conducted a systematic review of 81 studies on code search to understand research trends, analyze key components of code search tools, and classify existing tools based on their focus on supporting different search tasks. Their findings highlighted outstanding challenges in existing studies and provided a research roadmap for future code search research. We compare the differences of each survey to our study in detail in Section\ref{related}.

\section{Methodology}

\label{Methodology}
To conduct the survey, we followed the SEGRESS guidelines provided by Kitchenham et al.~\cite{kitchenham2022segress} to collect relevant papers published from 2012 to 2023.

\subsection{Selection Process}

\begin{figure*}[t!]
    \centering
    \includegraphics[width=\textwidth]{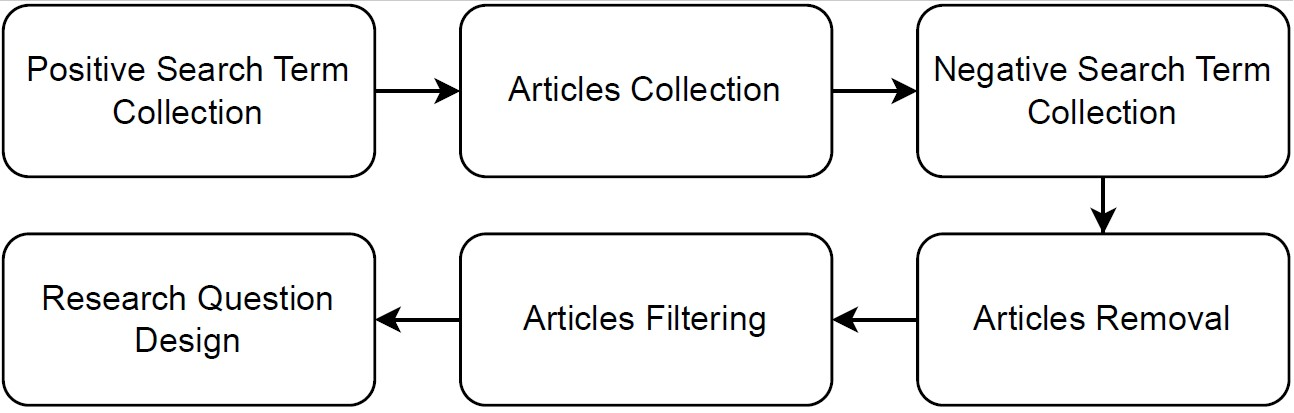}
     \caption{The selection process of collected research articles}
    \label{fig:process}
\end{figure*}


Figure \ref{fig:process} shows the overview of the article selection process. Initially, we identified positive search terms by conducting a manual investigation of several well-known API recommendation papers. Subsequently, we conducted online searches using these selected terms to gather relevant articles. During this phase, we observed that numerous irrelevant articles also matched the search terms. To address this issue, we conducted further investigations and compiled a list of negative search terms to exclude such articles. Following this, we engaged in discussions with the reviewer team to establish a set of filtering criteria. Finally, we reviewed each remaining paper to formulate the research questions. The following paragraph provides a detailed account of the selection process.

\subsection{Search Term Collection}

We initiated the search term collection process by manually reviewing the top 10 results (ranked by the number of citations) from Google Scholar using the term ``API recommendation'' to assess the relevance of each paper to the topic of API recommendation. After that, we manually extracted ten keywords from each of the selected papers as search terms. Then these keywords were used in another round of searches across various publication platforms, including ACM, IEEE Explore, Science Direct, and Google Scholar. We collected papers from these platforms that contain the selected keywords in their titles or abstracts. The selection of papers was based on the logical operator OR, meaning that a paper was included if it contained any of the keywords. For example, the search criteria utilized were ``API recommendation'' OR ``API method recommendation''. We chose not to split the search terms and apply OR logic on a word-by-word basis, as this approach tended to yield a large number of irrelevant papers. For instance, terms like ``API'', ``method'', and ``recommendation'' are generic, and searching for them individually would result in papers from diverse fields unrelated to API recommendation. After manually examining the papers, we discovered a few false-positive patterns. Therefore, we automatically eliminate tokens with irrelevant keywords. Table~\ref{table_1} shows the keywords we selected and removed for paper selection. 

\begin{table}[t!]
\caption{Key words for inclusion and exclusion}
\label{table_1}
\centering
{%
\begin{tabular}{cl}
\toprule
\textbf{Decision} & \textbf{Criteria} \\
\midrule
  \cmark &  API recommendation, API learning, API search, API retrieval\\
  \cmark & API method recommendation, API Q \& A, API question answering\\
\midrule
  \xmark & Program repair, API misuse, Malware, Bug fix, Web API \\
  \xmark & Error message, API migration, Vulnerability detection \\

\bottomrule
\end{tabular}%
}
\end{table}

\begin{table}[t!]
\caption{Criteria of inclusion and exclusion}
\label{table_1_1}
\renewcommand{\arraystretch}{0.1}
\centering
{%
\begin{tabular}{cl}
\toprule
\textbf{Decision} & \textbf{Criteria} \\
\midrule
  \cmark & Papers that are published in top-ranked conferences or journals. \\
  \cmark & The paper must propose a new tool with a complete \\
  & evaluation and discussion on its performance and impact. \\

\midrule
  \xmark & Papers that are not proposing new techniques\\
  & or do not have enough evaluation, including review papers,\\ 
  & workshop papers, short papers, keynotes, and posters. \\
  \xmark & The paper is an extended paper (we keep the initial version only). \\

\bottomrule
\end{tabular}%
}
\end{table}

\subsection{Paper Selection Criteria}

After the initial collection of papers, the next step is the establishment of a set of filtering criteria. For instance, the selected paper must be either a conference paper or a journal paper, and it must propose new tools. We have summarized a list of criteria that an accepted paper should meet, as well as a list of criteria that papers should not meet. Table \ref{table_1_1} presents the inclusion and exclusion criteria. After applying the aforementioned filters, all the papers were subjected to manual examination. We identified a total of 75 papers. Subsequently, we thoroughly reviewed the full content of these papers, resulting in a final selection of 34 papers.

\subsection{Data Extraction and Analysis}


We carefully reviewed the remaining 34 papers and extracted relevant data that could contribute to answering our proposed research questions. To ensure the effectiveness of this review process, we employed a systematic approach to standardize the process. Initially, we distributed the selected papers among our team of expert reviewers, with each reviewer assigned the task of thoroughly reading 5 papers. Subsequently, we conducted a discussion among the team members to identify common elements and content among these papers. We formulated our research questions based on these discussions.

To effectively conduct the data collection process, we designed a structured form for information collection. The form includes a list of questions such as: What are the publication venues? What is the data collection process? What is the modeling technique? What are the metrics for evaluation? Reviewers were instructed to complete this form while reading new papers. Our research questions focus on four primary categories: publication trends, data collection, modeling, and evaluation.







\subsection{Study Quality Assessment}

To better assess the fairness of our approach, we conducted a bias assessment. 
Specifically, we adopted a semi-automatic approach for completing the assessment form. Initially, the first three authors worked independently to extract relevant information from each paper and categorize it after that. Then, they manually verified the accuracy of these categorizations. 
All authors are involved in manually reviewing the forms completed by the first three authors and resolving any disagreements that occurred during the process to ensure consensus among reviewers.


\section{RQ1: What are the trends in API recommendation?}
\label{RQ1}

We extract information related to publications in the 34 API recommendation papers and analyze the trends in API recommendation studies.

\subsection{What is the distribution of publication time in API recommendation studies?}


\begin{figure*}[t!]
    \centering
    \begin{minipage}{.5\textwidth}
      \centering
      \includegraphics[width=\textwidth]{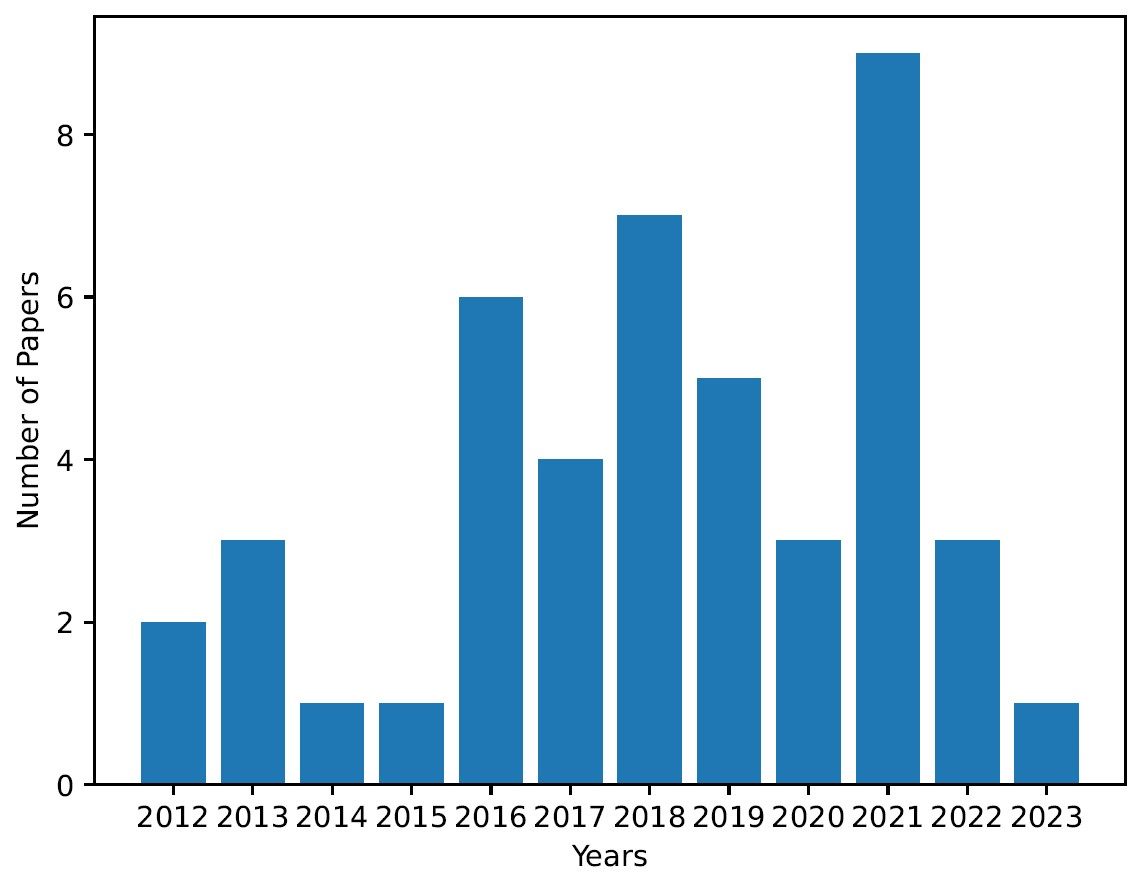}
      \captionof{figure}{The number of paper published each year}
      \label{fig:pub_year}
    \end{minipage}
\end{figure*}


Figure \ref{fig:pub_year} shows the annual quantity of API recommendation papers. It demonstrates an increasing interest in API recommendation over the past decade. Notably, 2016 and 2018 marked the first two peaks in the number of publications, coinciding with the release of {\RACK}~\cite{rahman2016rack} and BIKER~\cite{huang2018api}, two significant contributions to the field of API recommendation. Although there was a decrease in 2020, the overall trend from 2012 to 2021 indicates an increasing level of interest in this research area. The combined number of publications in 2016, 2018, and 2021 represents an additional 62\% of the total publications. To assess the publication time trend, we conducted a Cos Stuart trend test~\cite{cox1955some} at a 5\% significance level, resulting in a p-value of less than 0.01, confirming a significant upward trend. In 2022 and 2023, there was a notable decrease in API recommendations. We attribute this decline to the increasing prominence of source code models and large language models, which have led to improvements in complete code generation tasks. Consequently, API recommendation, as an intermediate sub-task, is not as popular as it once was.

\subsection{What is the distribution of publication venues in API recommendation studies?}

\begin{table}[t!]
\caption{Conference and journal publication count and venue names}
\label{table_9}
\renewcommand{\arraystretch}{0.3}
\resizebox{\columnwidth}{!}{%
\begin{tabular}{cccl}
\toprule
No & Acronym & Count& Full name\\
\midrule

1   & ICSE      & 5 & International Conference on Software Engineering\\
2   & ASE       & 5 & International Conference on Automated Software Engineering\\
3   & ECSE/FSE  & 3 & ACM SIGSOFT Symposium on the Foundation of Software Engineering \\
4  & IEEE Access& 2 & IEEE Access\\
5  & IST        & 2 & Journal of Information and Software Technology\\
6   & QRS       & 2 & International Conference on Software Quality, Reliability and Security\\
7   & TSE       & 2 & IEEE Transactions on Software Engineering\\
8   & SANER     & 2 & International Conference on Software Analysis, Evolution and Re-engineering\\
9  & APSEC     & 1 & Asia Pacific Software Engineering Conference\\
10   & CSMR     & 1 & European Conference on Software Maintenance and Reengineering\\
11   & EMNLP    & 1 & Empirical Methods in Natural Language Processing\\
12  & ICCSE     & 1 & International Conference of Computer Science and Engineering\\
13  & ICCEE     & 1 & International Conference on Computer and Electrical Engineering\\
14  & IET       & 1 & The Institution of Engineering and Technology\\
15  & SCP       & 1 & Journal of Science of Computer Programming\\
16  & TOSEM     & 1 & ACM Transactions on Software Engineering and Methodology\\
17  & JCST      & 1 & Journal of Computer Science and Technology\\
18   & QRS-C    & 1 & International Conference on Software Quality, Reliability and Security Companion \\
19   & MSR    & 1 & International Conference on Mining Software Repositories \\

\bottomrule
\end{tabular}}
\end{table}
The publications are organized by conference and journal names in Table \ref{table_9}. 
Among the 20 publication venues, the top three venues with the most publications for API recommendation papers are ICSE, FSE, and ASE, which are also A+ conferences in the CORE ranking system. The number of API recommendation papers published in the top three venues accounts for 45\% of the total number of API recommendation publications.

\subsection{what are the frequent key directions and concepts in API recommendation studies?}

\begin{figure*}[t!]
    \centering
    \begin{minipage}{.5\textwidth}
      \centering
      \includegraphics[width=\textwidth]{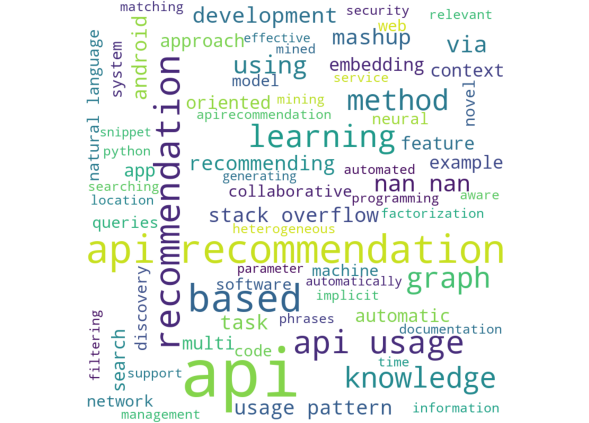}
      \captionof{figure}{The word cloud of collected papers}
      \label{fig:cloud}
    \end{minipage}%
    \begin{minipage}{.5\textwidth}
      \centering
      \includegraphics[width=\textwidth]{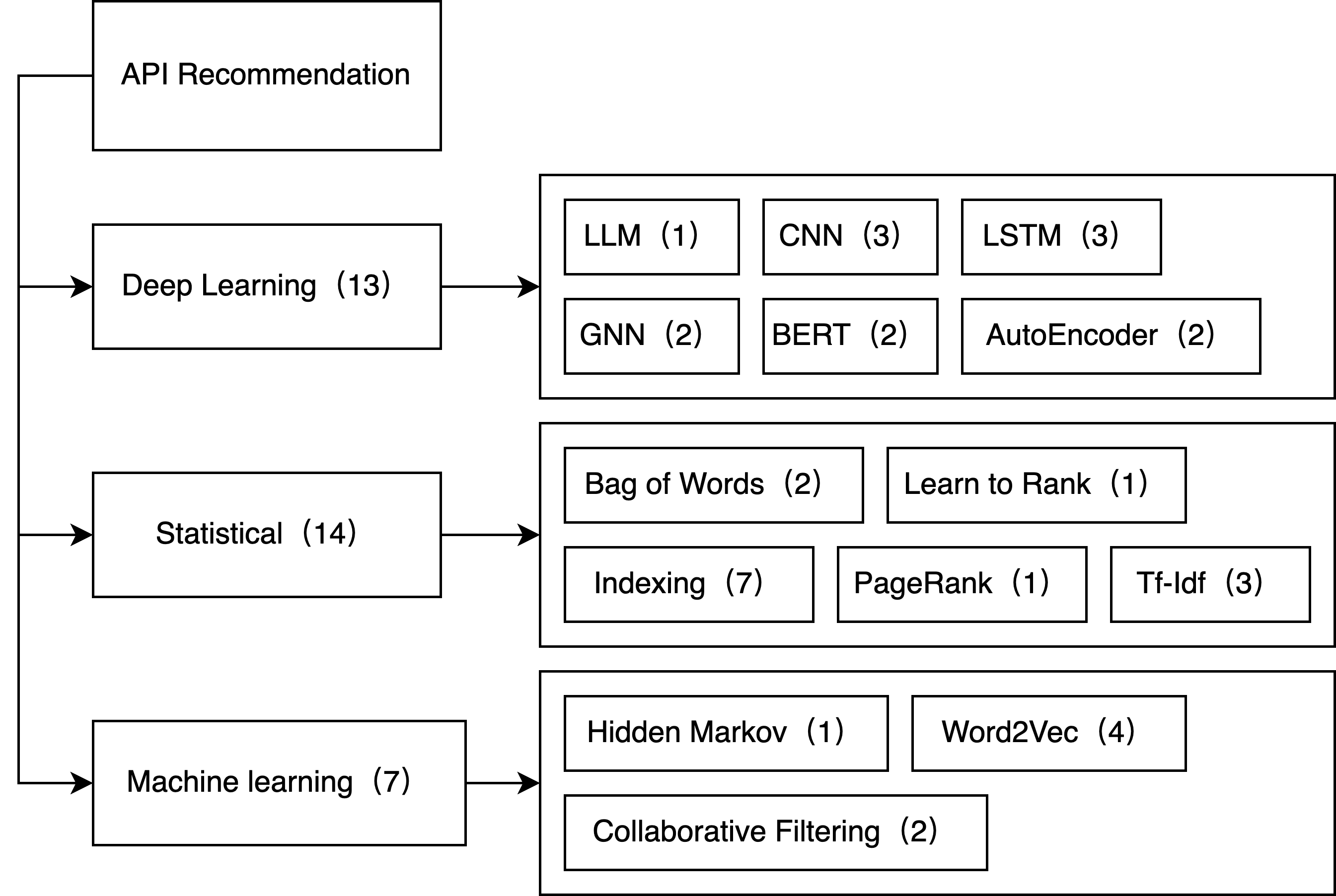}
      \captionof{figure}{Distribution of API recommendation tools}
      \label{fig:directions}
    \end{minipage}
\end{figure*}

To better understand the publication trends on the directions and concept of API recommendation research field. We performed Keyword frequency analysis and presented the result as a word cloud to visualize the result. Figure \ref{fig:cloud} shows a word cloud chart for the keyword in the titles. We created a word cloud based on the keywords in the titles to provide an intuitive visualization of trends and keywords. We tokenized the titles and generated the word cloud using the Python word cloud package. From the figure, we can see that the common words in API recommendation are API usage, knowledge, and learning. To gain a deeper understanding of the current direction of API recommendation research, we carefully read through all the collected articles and categorized the papers. Since all the papers are closely related to query-based API recommendations, the input format, output format, and application directions of these papers are almost identical. It is less meaningful to categorize them based on the above factors. We compared multiple factors and believe that the most meaningful way to categorize the papers is based on the key differences in the modeling algorithms they apply. Figure \ref{fig:directions} shows the categorization of all papers.

Researchers generally consider API recommendation as a retrieval task rather than a generation task. In general, there are three primary directions in the modeling techniques of these tools. In the early years of API recommendation, researchers approached the API recommendation problem using statistical approaches such as Indexing~\cite{tran2013api,xie2020api,heinemann2012identifier,rahman2016rack,nguyen2012grapacc,nguyen2016api,raghothaman2016swim}, Tf-idf~\cite{gao2021api,yuan2019api,zhang2017recommending,ling2019graph}, bag-of-words (BOW)~\cite{thung2013automatic}, PageRank~\cite{ponzanelli2017supporting} and Learn-to-rank~\cite{zhou2021boosting}. Starting in 2016, the improvements in machine learning techniques drew researcher's attention. Researchers have started applying the latest machine learning techniques such as word2Vec~\cite{liu2019design,nguyen2017exploring,lee2021mining,van2017combining}, Collaborative filtering~\cite{wang2021hybrid,nguyen2021recommending}, and Hidden Markov Model~\cite{nguyen2016learning} to API recommendation tasks. Compared to manually designed scoring functions, machine-learning techniques such as word2Vec bring more flexibility to the ranking and improvement in overall performance.

In recent years, researchers have started applying deep-learning techniques such as BERT~\cite{yin2021api,wei2022clear,li2020fusion}, LSTM~\cite{tian2018automatically,yan2018learning,gu2016deep,gu2018deep}, GNN~\cite{chen2021holistic,chen2021novel}, and CNN~\cite{sworna2023apiro,sun2019enabling,wang2021plot2api}. Deep learning techniques enabled more possibility for API recommendation tasks as they allow advanced semantic parsing that clusters the intention of query based on query semantics without explicit design. Also, Generative models such as LLM~\cite{kang2021apirecx} allows user to query APIs in the form of open question-answering by only giving API documents as a knowledge base context in the prompt.

\mybox{Answer to RQ1}{red!40}{gray!10}{
The overall trend from 2012 to 2021 indicates a growing level of interest in API recommendations. The three most frequent keywords in API recommendation papers are API usage, knowledge, and learning. Based on our categorization, the three primary techniques applied in API recommendation are deep learning, statistical-based approach, and machine learning.
}

\section{RQ2: What are the common data sources in API recommendations?}
\label{RQ2}


\begin{table}[t]
\caption{Common data sources}
\label{table_10}
\begin{tabular}{ccccccc}
\toprule


Data source           & Code Context & Code Completeness & Query  & Quality & Quantity  & Count  \\
\midrule
Github                & \cmark       & High              & \xmark & Medium & High       & 16     \\
Stack Overflow        & \xmark       & Low               & \cmark & Low    & Medium     & 10     \\
Online Document       & \xmark       & High              & \xmark & High   & Low        & 8      \\

\bottomrule        
\end{tabular}
\end{table}

Selecting an appropriate dataset is a crucial step in developing an API recommendation system, which is also the first step in creating a tool that can recommend APIs. The choice of data source has a significant impact on the scope, potential use cases, and applicability of the API recommendation. The dataset can be sourced from various origins for different application areas of the API recommendation. 
Table \ref{table_10} shows the source of the dataset of 34 collected articles. Specifically, GitHub and Stack Overflow are the two largest data sources, accounting for 76.5\% of the total number of papers. We have analyzed three data sources across five dimensions, namely code context, code completeness, presence of data queries, quality of data, and quantity of data. Each data source has its own strengths and weaknesses. 

GitHub is one of the biggest open-source data repositories in the world, with over 128 million repositories. Out of these repositories, around 258,000 are Python repositories, and around 240,000 are Java repositories. Researchers prefer to collect data from many repositories rather than just one or a few. This is because GitHub hosts a variety of application directions, and existing works aim to develop general API recommendation tools that can be applied across multiple projects instead of project-specific API recommendations. It is worth noting that researchers prefer to collect GitHub datasets themselves instead of reusing existing datasets~\cite{kalliamvakou2014promises}. This is because the API data on GitHub is time-sensitive. GitHub projects are updated frequently, from daily to monthly, and the code within these projects also evolves. As a result, collecting data at different points in time can yield significantly different datasets. Additionally, some repositories may disappear over time due to deprecation or lack of maintenance. To ensure that recommendation tools remain valid for the most recent data, most studies re-collect the most up-to-date data. Although the dataset is different from each other, most studies employed similar collection methods to those of previously published research. When comparing the GitHub data source with the other two data sources, it becomes clear that GitHub has the advantage of having complete code context, making it more convenient for code analysis. However, one common issue with this dataset is the lack of coding queries. All the queries that are paired with GitHub have to be generated or inferred from code comments. In summary, GitHub provides comprehensive source code information, including code context, complete code snippets, and a high number of samples. However, Github lacks user queries, and all queries paired with APIs must be inferred or transformed from source code comments, which can cause difficulties in generating query-API pairs~\cite{gu2016deep,gu2018deep}.

Stack Overflow is a popular question-and-answer forum for software developers. The data on Stack Overflow is presented as posts, which consist of a question title, question description, an accepted answer and additional answers. Researchers typically structure the data obtained from Stack Overflow as question-answer pairs and use natural language processing techniques to extract features from the questions. They then use mathematical modeling to establish the relationship between the questions and their corresponding code. All the datasets obtained from Stack Overflow consist of question-answer pairs where the question component includes an English sentence describing an API-related question. The answer component usually contains a code snippet, which either resolves the problem or includes an API method that could potentially resolve the issue. However, it's important to note that the quality of the data varies greatly since the level of experience of each contributor varies from beginner to expert. In summary, Stack Overflow (SO) fulfills the missing information of user queries, as the query can be directly collected from the title of SO posts. However, since SO is a forum for open discussion, the quality of answers varies. The answers usually contain discussions and conversations, and the completeness of code snippets is not guaranteed. As a result, filtering and post-processing tasks must be applied to SO data to improve its quality~\cite{huang2018api,wei2022clear,rahman2016rack}. 

 Online documents are also valuable sources of data, in addition to Stack Overflow and GitHub. This term refers to online materials created by community members, and researchers often collect data from such documents as a complement to other datasets. Online documents, provided by official websites or online tutorials, are regarded as the most reliable source of API feature and description information since they are manually written and reviewed by developers. However, writing documents is time-consuming, so the amount of samples from this data source is low. Examples of such online documents include official API documents, GitHub Issues, Jira Tickets, and online tutorials. Since these Q\&A pairs are reviewed manually and the amount is relatively small compared to data scraped from GitHub and Stack Overflow, they are typically considered high-quality and they are sometimes used as a reference for model evaluation~\cite{wei2022clear,chen2021holistic,feng2020codebert,allamanis2018survey}.

\mybox{Answer to RQ2}{red!40}{gray!10}{
\small{ Over the last 10 years, GitHub and Stack Overflow have emerged as the two largest data sources for API recommendation studies, collectively accounting for 61.2\% of the total number of papers. Additionally, online document collected from various sources is also one of the common choices.}
}





\section{RQ3: What are common data processing techniques applied in the API recommendation tools?}
\label{RQ3}

\subsection{What are the common API Usage Pattern Types?}
The API extraction techniques can be categorized into two distinct categories based on the data source: single-API usage information and multi-API usage information. The latter is often referred to as API usage patterns, which are sequences of frequent API method calls extracted from software source code.

\subsubsection{Single API Recommendation Task}
For single API recommendation tasks, when building the API recommendation knowledge base, researchers often gather information related to individual API usage, including API usage examples and API method descriptions. For instance, Xie et al.~\cite{xie2020api} proposed API method recommendation techniques based on the relationship between API method names and their descriptions. Since the primary focus of the study is the analysis and reformulation of natural language descriptions of API methods, it suffices to collect only the API descriptions for knowledge base construction.

\subsubsection{Multi-API Recommendation Task}
For multi-API recommendation tasks or API sequence recommendation tasks, it is essential to conduct API pattern usage mining to extract and store frequently used API sequences in the knowledge base before constructing the model. As part of the API recommendation knowledge base, API usage patterns are collected during the model construction process. These patterns complement the information gathered for single-API usage. In certain studies~\cite{monperrus2016searching,lee2018comment,chan2012searching}, the API recommendation tool also provides code snippets as API usage examples. In these cases, the tool gathers mapping information between code fragments and API method names or API usage patterns.

\subsection{What are the common API usage extraction methods?}
\label{RQ3_1}
\begin{table}[t!]
\caption{Common API usage extraction methods}
\label{table_23}
\renewcommand{\arraystretch}{0.1}
\begin{tabular}{ccccc}
\toprule
No & Extraction Method & Github & StackOverflow & Online Document      \\
\midrule

1   & Graph Analysis   & \cite{nguyen2012grapacc,chen2021novel,liu2018effective} & - & \cite{chen2021holistic,
ling2019graph,
nguyen2015recommending}\\
2   & AST Parsing       & \cite{yan2018learning,nguyen2016api,heinemann2012identifier}&\cite{wang2021plot2api}  &\cite{sun2019enabling}\\
3   & Code Parsing   &\begin{tabular}{@{}c@{}}\cite{kang2021apirecx,liu2019autoencoder,liu2019design,nguyen2017exploring} \\ \cite{raghothaman2016swim,tian2018automatically,wang2021hybrid,zhao2021icon2code}\end{tabular}& - & - \\
4   & Document Parsing & \cite{gu2016deep,
xie2020api}& 
\begin{tabular}{@{}c@{}}\cite{huang2018api,irsan2023picaso,lee2021mining,li2020fusion,rahman2016rack} \\ \cite{yuan2019api,zhang2017recommending,zhou2021boosting,wei2022clear}\end{tabular}

&\cite{gao2021api,
ponzanelli2017supporting,
sworna2023apiro,
thung2013automatic}\\

\bottomrule
\end{tabular}
\end{table}



API usage extraction plays a crucial role in building API recommendation tools. Within our collection of papers, various model architectures and technical approaches have resulted in a range of API usage pattern collection techniques. We have gathered, identified, and classified the API collection techniques used in our paper collection. To build the model, an API knowledge base is essential for recommending API methods. Researchers employ various data mining techniques to extract API usage data for constructing this knowledge base. The API knowledge base gathers API usage information from sources like API documentation and public source code archives. 

Table \ref{table_23} presents the API usage extraction methods employed by existing studies. The rows are the method employed and the columns are the data source. The graph clearly illustrates that document parsing and code parsing are the most popular tools. The following paragraph describes the data representations for API recommendation studies, including API usage extraction types, such as single API recommendation and multiple API recommendation. It also includes the types of data parsing, including document parsing and source code parsing.

\subsubsection{Graph Analysis}
Several studies have utilized graph analysis for API usage pattern extraction~\cite{nguyen2012grapacc, nguyen2015recommending, chen2021novel, liu2018effective}. Nguyen et al.~\cite{nguyen2012grapacc} proposed GraPacc for context-sensitive API recommendation based on API usage patterns. They extract API usage patterns from API knowledge graphs. Nguyen et al.~\cite{nguyen2015recommending} proposed HAPI, a statistical generative model of API usages based on Hidden Markov Model. They build the API usage representations using an API call graph model. Chen et al.~\cite{chen2021novel} proposed JARST for recommending API usages by analyzing the structure information and text information of the source code. They extract API usage patterns from structured source code using graph analysis techniques.



\subsubsection{AST Parsering}
An Abstract Syntax Tree (AST) represents source code as a tree structure. To convert source code into an AST structure, the code must have a formal definition in a formal language, adhering to a set of rules defined by formal grammar~\cite{jones2003abstract}. AST representation is one of the most crucial static analysis techniques employed in software engineering. Multiple research fields, including code clone detection~\cite{baxter1998clone}, defect prediction~\cite{cai2019abstract}, vulnerability detection~\cite{yamaguchi2012generalized}, and automated bug repair~\cite{koyuncu2020fixminer}, have adopted the AST data representation technique. 

In API recommendation, only a few studies apply AST parsers~\cite{he2021pyart, yan2018learning, wang2021plot2api}.
He et al.~\cite{he2021pyart} proposed PyART for recommending APIs for Python programs in real-time. They extract API usage patterns by analyzing the AST nodes of Python code. Yan et al.~\cite{yan2018learning} proposed APIHelper, which predicts API sequence patterns using a Long Short-Term Memory (LSTM) network. They extract API usage pattern sequences using a Java AST analyzer. Wang et al.~\cite{wang2021plot2api} proposed plot2api, which recommends APIs based on chart plots. They extract the ground truth APIs using an AST parser.

\subsubsection{Code and Document Parsing}

Many existing studies extract API usage information directly from API documentation~\cite{nguyen2012grapacc,nguyen2015recommending,chen2021novel,liu2018effective}, primarily sourced from official documentation websites such as the Java JDK official document website, Android Developer guides, and Python library websites.

\paragraph{Java JDK}
The official Java JDK API documentation serves as the most frequently cited data source for API usage in this category. The format of the official Java JDK API documentation is highly structured and consistent, allowing for automated extraction of API information through keyword matching. However, the level of detail extracted from Java JDK official documents regarding API usage can vary among tools.
For example, consider the API method "java.lang.Iterable.forEach{()}"\footnote{Java for each documentation page: https://docs.oracle.com/en/java/javase/11/docs/api/java.base/java/lang/Iterable.html}, which belongs to the Java SE standard library version 11. This method serves as a syntactic alternative to the for-loop syntax. Its function description is as follows: \textit{``Performs the given action for each element of the Iterable until all elements have been processed or the action throws an exception''}. In the documentation, the API name is enclosed within {<pre>} tags with the keyword ``methodSignature,'' and the API description resides in a {<div>} block adjacent to the {<pre>} block. Additionally, API usage examples for the method ``java.lang.Iterable.forEach()'' are provided within the {<pre>} tags inside the API description. The entire API documentation website adheres to the same formatting rules. Researchers can extract API usage information using regular expressions or hypertext parsing tools such as BeautifulSoup.

Alternatively, one can extract API usage information from the Java library by parsing comments within the codebase using the JavaDoc tool. JavaDoc follows a standardized format for documenting API method descriptions within the codebase. Java docstrings always precede the method definition and consistently begin with "{/**}" and end with "{*/}." Researchers can employ the JavaDoc tools to extract API documentation from the codebase of any Java library that adheres to this Java docstring standard.

\paragraph{Android Developers Guide}
Several studies have focused on Android SDK API methods, as documented in ~\cite{liu2019autoencoder,nguyen2021recommending,gao2021api,sun2021task}. Android, an operating system built on top of the Linux kernel, is primarily designed for tablets and mobile devices. The official Android SDK API reference website, known as the Android developer guide\footnote{https://developer.android.com/reference/com/google/android/play/core/install/model/ActivityResult.html}, provides dedicated pages for each API class, offering comprehensive details about the APIs. For example, the android.play.core.install.model.ActivityResult class is one of the core APIs in Android. In the detailed section of the webpage, the API description states, \textit{Custom Activity results for the in-app update flow.} Additionally, in the \textit{inherited} section, all inherited APIs are listed with their descriptions. Researchers can extract this information using regular expressions or hypertext parsing tools.

\paragraph{Python Standard Library}
There is a growing focus on the study of Python language API method documentation, as demonstrated by prior works such as ~\cite{richardson2017function,yin2017syntactic,svyatkovskiy2019pythia,wang2021plot2api,d2016collective,he2021pyart}. Python3, released in 2008, has rapidly emerged as one of the most prominent programming languages, finding applications in diverse fields such as machine learning, deep learning, web scraping, and automation. Researchers commonly regard the Python Standard Library as the authoritative source for Python API documentation\footnote{https://docs.python.org/3/library/}. The API documentation website follows a structured format, with a dedicated page for each API class. These pages typically include details about inherited methods, usage descriptions, and occasionally, API usage examples. For instance, consider the re.split() function, which is a commonly used function in the Python regular expression module. On the documentation page for the re API, the re.split() function is enclosed within a {<code>} tag with the class name {sig-name,} indicating the function's signature name. Beneath the method name, the description of re.split() reads as follows: \textit{Split a string based on occurrences of the pattern.} If capturing parentheses are employed in the pattern, the text of all groups in the pattern is also included in the resulting list\footnote{https://docs.python.org/3/library/re.html}. Apart from the standard Python API documentation, there are third-party library documentation websites. Nevertheless, these websites often have distinct designs and layouts compared to the standard Python API library. Extracting API information from these websites typically necessitates the use of customized web scraping scripts. 
pyDoc is the documentation tool for Python's codebase, akin to JavaDoc. A pyDoc docstring should be positioned within a method's definition, and its format commences and concludes with triple quotation marks. If the docstring of a Python library adheres to the pyDoc standard, the pyDoc tool can be employed to extract API usage information from any Python library.

\paragraph{QA Website and Online Tutorials}

API recommendation tools can provide API methods in response to natural language queries, utilizing data collected from Online Question Answering (QA) websites and online tutorials~\cite{rahman2016rack, ye2016word, sun2021task, huang2018api}. Rahman et al.~\cite{rahman2016rack} establish a relationship mapping between API usage information and related programming questions by extracting knowledge from the Stack Overflow Q\&A site. This approach aims to bridge the API knowledge gap that often exists between natural language questions and API usage. 
Sun et al.~\cite{sun2019enabling} create an API knowledge graph by extracting API usage patterns from Stack Overflow, enhancing the programming learning experience. Huang et al.~\cite{huang2018api} combine API documentation and programming questions from Stack Overflow to improve API recommendation ranking. Ye et al.~\cite{ye2016word} extract programming questions from Stack Overflow and utilize API documents for word similarity evaluation, specifically tailored to software engineering information retrieval tasks.

\subsection{Other Tools}

Considering the variety of API-related Data sources, researchers apply a mixture of multiple methods for API usage extractions. There are many other tools used for API usage information extraction.
For instance, He et al.~\cite{he2021pyart} employ data-flow analysis for real-time API recommendation. They encode not only the API method call sequence but also information about how data is passed between objects, as the data flow may contain crucial information for the recommendation system's decision-making.
Gao et al.~\cite{gao2021api} utilize APKtool to extract the resource folder and XML files from the standard Android Application distribution file (APK). From these XML files, they extract the UI element API method names and API usage patterns. Raghothaman et al.~\cite{raghothaman2016swim} utilize the standard C\# AST parsing tool, Roslyn C\# AST parser. The Roslyn AST parser is used in a manner comparable to the Python AST parser and the Java AST parser. 
Additionally, several studies employ static analysis techniques without specifying the precise tool used for API usage pattern extraction~\cite{nguyen2021recommending, wang2021hybrid, asaduzzaman2017recommending}.

\mybox{Answer to RQ3}{red!40}{gray!10}{
\small{
Researchers extract text-based API usage information primarily from official API documentation sources such as the Java JDK Website, Android developer guides, and the Python Standard Library, with each source offering structured and consistent formats for information extraction. The API extraction techniques can be categorized into two distinct categories based on the data source: single-API usage information and multi-API usage information. Out of the four API usage methods, document parsing and code parsing are the most popular methods for data extraction.}
}

\section{RQ4: What are the common modeling techniques in API recommendation tools?}
\label{RQ4}
The most crucial aspect of API recommendation is the modeling process, which also plays a significant role in differentiating one paper from another in terms of originality.

In this section, we categorize the models used in our collection of papers into several categories and subcategories, providing a detailed description of each subcategory. The three primary modeling directions are statistical approach, machine learning, and deep learning. we distinguish the deep learning approach from the other two based on wither the layer of the model is more than three~\cite{lecun2015deep}. It's expected that a significant proportion of machine learning and deep learning models will be applied to API recommendation. This expectation comes from considering the API recommendation research field as a sub-topic of the recommendation engine research field, where machine learning approaches are among the most popular and effective. 

\subsection{Statistic-based Approach}
\begin{table}[t!]
\caption{Common statistical modeling techniques }
\label{table_14}
\renewcommand{\arraystretch}{0.3}
\begin{tabular}{cll}
\toprule
No & Model &  Reference      \\
\midrule



1 & Indexing & \cite{nguyen2012grapacc,
liu2018effective,
raghothaman2016swim,
rahman2016rack,
heinemann2012identifier,
xie2020api,
nguyen2016api}\\
2 & Bag-of-Words   &\cite{ling2019graph,
thung2013automatic}\\
3 & TF-IDF         &\cite{gao2021api,
yuan2019api,
zhang2017recommending}\\
4 & PageRank       &\cite{ponzanelli2017supporting}\\
5 & Learn to Rank  &\cite{zhou2021boosting}\\

\bottomrule
\end{tabular}
\end{table}

The statistic-based models employed by API recommendation tools are listed in Table \ref{table_14}. We categorize these algorithms based on statistics into five categories. Among these approaches, the most popular model is the Indexing model. This section describes how researchers adapt each statistical model to their respective API recommendation tool.

\subsubsection{Indexing model}

The most prevalent text processing technique is indexing~\cite{tran2013api,xie2020api,heinemann2012identifier,rahman2016rack,nguyen2012grapacc,nguyen2016api,raghothaman2016swim}. The indexing technique identifies the correlation between the keywords and the APIs and combines them together to a look-up table. For a new query, the system matches the keywords in the user query to related APIs. For example, Rahman et al.~\cite{rahman2016rack} designed a scoring mechanism and built a keywords-API table based on the connection between keywords in the StackOverflow title and the APIs in the post. Nguyen et al.~\cite{nguyen2016api} constructed an indexing system to capture the fine-grained information from the code changes and recommend APIs based on the keywords in a new query.

\subsubsection{Term Frequency-Inverse Document Frequency(TF-IDF)}
The other prevalent modeling technique is TF-IDF, an algorithm for text vectorization that converts each token in a sequence into a numeric value by calculating its frequency statistics. Many researchers in the API recommendation field employ TF-IDF as a fundamental text processing and source code processing technique because TF-IDF can be applied to general tokenizable sequences, including source code.

For example, Zhang et al. ~\cite{zhang2017recommending} use the TF-IDF algorithm to transform API-related questions and functional descriptions into vectors. Gao et al. ~\cite{gao2021api} utilize TF-IDF to determine the significance of APIs within their customized API set. Thung et al. ~\cite{thung2013automatic} employ the TF-IDF algorithm to convert sequences into vectors for vector space modeling. According to our study, API recommendation researchers primarily use TF-IDF for text or code vectorization and for calculating the importance score of API method calls.

\subsubsection{Bag-of-words, PageRank, and Learn to Rank}

Although most researchers approach the API recommendation task by manually designing a variety of scoring functions, there are still a few statistical approaches other than TF-IDF and Indexing. Ling et al.~\cite{ling2019graph} apply a mixture of API call graph analysis and Bag-of-Words algorithm to get the graph embeddings of the API sequence. Ponzanelli et al.~\cite{ponzanelli2017supporting} proposed an Extended version of the PageRank Algorithm to recommend and rerank the content collected from online API-related sources with consideration of prominence and complementarity. Zhou et al.~\cite{zhou2021boosting} apply the Learning-to-rank (LTR) model to continuously improve the performance of API recommendation by leveraging the API-related information feature as the training data and feedback feature as the feedback of the performance improvement signal.

\subsection{Machine Learning-based Approach}
\begin{table}[t!]
\caption{Common machine learning modeling technique }
\label{table_14}
\renewcommand{\arraystretch}{0.3}
\begin{tabular}{cll}
\toprule
No & model &  Reference      \\
\midrule

1 & Word2Vec & \cite{huang2018api,
liu2019design,
lee2021mining,
nguyen2017exploring}\\
2 & Collaborative Filtering &\cite{wang2021hybrid,
zhao2021icon2code}\\
3 & Hidden Markov Model &\cite{nguyen2015recommending}\\
\bottomrule
\end{tabular}
\end{table}



\subsubsection{Word2Vec}
Word2vec is a neural network natural language processing model proposed by Mikolov, Tomas, et al. ~\cite{mikolov2013efficient} in 2013, which converts word tokens into vectors. It discovers word vectors by analyzing text sequences from a large text corpus. Initially, it was used to identify synonyms because words with similar meanings exhibit similar usage patterns in the corpus. Since similar API method calls also have comparable usage patterns, the Word2vec model can be extended to API method call recommendations in the API recommendation field. Doc2vec~\cite{le2014distributed} is an extension of the Word2vec model, designed to convert documents into vectors. Huang et al. ~\cite{huang2018api} proposed BIKER, a Word2vec-based API recommendation model. It models API method calls and descriptions and then recommends API method calls by comparing API method call vectors for similarity. As API recommendation research advanced, Word2vec became a benchmark for source code and language modeling.

\subsubsection{Collaborative Filtering}
Collaborative Filtering is a common recommendation engine technique that has been applied to a variety of recommendation tasks, including those involving movies, music, books, and online shopping. In a collaborative filtering system, there are two types of objects: items and users. The CF system constructs a matrix based on the relationship between items and users and then ranks the vector similarity between the query item and the items in the matrix to make recommendations. There are two main types of CF techniques: Memory-based collaborative filtering and Model-based collaborative filtering.

Memory-based collaborative filtering is based on the assumption that similar users have similar preferences and, consequently, similar tastes. This approach is used to predict item ratings. Typically, the ratings in the training dataset are normalized to address the issue of rating bias. Rating bias is the observation that some users tend to give relatively low scores for all the items they review, while others tend to give high scores. To address this issue, the ratings should be normalized based on each user's rating distribution.

In API recommendation tasks, memory-based CF often encounters the sparse matrix problem due to the large number of API methods in a dataset and the fact that each code snippet typically contains only a few API method calls, leaving the rest of the vector empty.

On the other hand, model-based collaborative filtering obtains item vectors through trained models. It is more compact, quicker to train, and less prone to overfitting than memory-based CF. For the API recommendation task, Wang et al. ~\cite{wang2021hybrid} used a hybrid item collaborative filtering technique. In their model, API method calls represent items, while API method declarations represent users. Memory-based collaborative filtering and model-based collaborative filtering are combined in the hybrid collaborative filtering system. Using memory-based CF, they first identify similar projects and method declarations at the project level. Then, they use model-based CF to complete the matrix. Finally, the completed API list is used to rank the candidates for API methods.

Nguyen et al. ~\cite{nguyen2021recommending} proposed FOCUS, a context-aware collaborative filtering approach for extracting API usage patterns from open-source software projects and making API recommendations based on API method usage similarity. They reformulate the API recommendation problem as a collaborative filtering problem between API usage patterns as items and projects as users.

\subsubsection{Hidden Markov Model}

The Hidden Markov Model (HMM) is a state-based model trained through a Markov process. The training objective of a Markov model is to uncover a hidden state model by observing a training dataset. Although the hidden state of a Markov model is not directly observable, it can be updated using observable data for model training. Nguyen et al. ~\cite{nguyen2016learning} introduced HAPI, a Hidden Markov Model (HMM) designed to model one or multiple API method calls. Given a HAPI state, the HMM model generates a method call and then transitions probabilistically to the next state. A custom algorithm is employed to train HAPI using API usage association information extracted from the SALAD ~\cite{nguyen2016learning} dataset.

\subsection{Deep Learning}
\begin{table}[t!]
\caption{Common deep learning modeling technique }
\label{table_15}
\renewcommand{\arraystretch}{0.3}
\begin{tabular}{cll}
\toprule
No & model &  Reference      \\
\midrule

1   & LSTM  & \cite{gu2016deep,
yan2018learning,
li2020fusion}\\
2   & BERT  &\cite{irsan2023picaso,
wei2022clear}\\
3   & CNN  & \cite{sworna2023apiro,
wang2021plot2api,
sun2019enabling}\\
4   & AutoEncoder  & \cite{tian2018automatically,
liu2019autoencoder}\\
5   & GNN  & \cite{chen2021holistic,
chen2021novel}\\
6   & LLM  & \cite{kang2021apirecx}\\



\bottomrule
\end{tabular}
\end{table}

Since the success of Deep learning-based applications such as AlphaGo~\cite{silver2016mastering}, Convectional neural network~\cite{o2015introduction} (CNN)-based image recognition, and Transformer-based language model, there has been considerable interest in the field ~\cite{zhang2020pegasus,raffel2020exploring,lewis2019bart,raffel2020exploring,lan2019albert,clark2020electra,liu2019roberta,yang2019xlnet,devlin2018bert,radford2018improving,vaswani2017attention}. Deep learning has become a popular and effective technique in a variety of research fields. Deep learning has also dramatically altered the paradigm of API recommendation research.

In this section, we introduced API recommendation tools that employ deep learning models, as well as how these tools employ deep learning models. Among these deep learning models, LSTM  GNN, and BERT are the most popular models for API recommendation.

\subsubsection{Long Short-Term Memory (LSTM)}

LSTM model is a type of recurrent deep learning model. In a recurrent model, the term "recurrent" signifies that the output of a recurrent layer is looped back to its input. The key distinction between the recurrent model and other deep learning models is that the recurrent model can handle an infinite amount of input data thanks to the propagation of the recurrent loop. A recurrent neural network propagates not only in space but also in the time dimension. In API recommendation research, several studies have employed LSTM as their recurrent model~\cite{tian2018automatically,yan2018learning,gu2016deep,gu2018deep}.

In 2016, Gu et al. ~\cite{gu2016deep} proposed DeepAPI, an RNN-based sequential model for recommending API method calls. It employs an encoder-decoder structure that converts the text sequence representing the programming question into a list of API method calls. The encoder is a recurrent RNN neural network that transforms the concealed state of each input token into a compressed concealed state C. The decoder then uses another RNN neural network to expand the compressed hidden state C into a list of API method hidden states, which it then maps to a list of API method invocations. Gu et al. ~\cite{gu2018deep} introduced DeepCodeSearch (DeepCS) in late 2018, which can be seen as an expansion of DeepAPI. Within the encoder-decoder architecture, DeepCS employs a parallel architecture that models source code and description concurrently. Compared to DeepAPI, this model provides additional information and improves the ranking method from probability to cosine similarity.



\subsubsection{Auto Encoder}

An autoencoder is a self-supervised model designed to discover dense representations of input data. During training, the objective of an autoencoder is to replicate the input data in the output using a neural network. The middle hidden layer of a trained autoencoder is considered the dense representation of the input data for a given set of input data. SADE is an autoencoder-based API recommendation system for Android programming proposed by Liu et al.~\cite{liu2019autoencoder}. The SADE system suggests API methods with usage patterns similar to a given code snippet. The system extracts information about API usage patterns from the import of API tools and then trains an autoencoder to learn the dense representation of API methods. The trained model serves as an API method input feature extraction model. Given input data, the model generates a ranked list of recommended API method calls.

\subsubsection{GNN}

A Graph Neural Network (GNN) ~\cite{scarselli2008graph} is a generalization of a neural network for processing graph data. Neural networks are composed of nodes and edges, which can be seen as graph structures. A Convolutional Neural Network (CNN) can also be considered a graph neural network, where the graph data forms a pixel grid. Similarly, in the context of a graph neural network, Recurrent Neural Networks (RNN) can be viewed as a chain structure, where each node represents an input token.
In the field of API recommendation, GNN is employed for modeling API method call graphs ~\cite{ling2021graph, chen2021novel, gu2019codekernel, chen2021holistic}. Using GNN for modeling call graphs is essential because API method calls are not always sequential, especially when considering execution flow. In scenarios involving concurrency, for instance, certain API call executions do not depend on the completion of a previous API call, allowing two APIs to execute in parallel. The actual execution of API calls forms a graph at runtime. In such cases, GNN is well-suited for processing API call graphs.

\subsubsection{Language Models}

A language model is a neural network model that learns the language based on the probability patterns of word sequences in the training corpus, similar to the Word2Vec model. Unlike Word2Vec, which focuses on learning synonyms, a language model assigns probabilities to input sequences. Language models find applications in tasks such as sentence vectorization and named entity recognition (NER).

Language models represent the cutting edge of source code modeling techniques. For instance, Yin et al.~\cite{yin2021api} employ the BERT model for the NER task of constructing an API method knowledge graph. The model identifies named entities and relationships from programming questions on Stack Overflow and then uses the extracted entities and relationships to construct the API knowledge graph. Kang et al.~\cite{kang2021apirecx} propose APIRecX, an API recommendation model based on the Generative Pre-Training (GPT) model. It models GitHub source code fragments and generates an API call list using beam search.

Feng et al.~\cite{feng2020codebert} introduce the CodeBERT model for general source code modeling, inspired by BERT~\cite{devlin2018bert}. CodeBERT's training objective is to enable the translation from natural language to programming language by achieving various training objectives, including masked language modeling~\cite{devlin2018bert} and replaced token detection. CodeBERT is trained using a substantial corpus of natural language to programming language data, making it versatile for various downstream tasks, including API recommendation and natural language code search.

Wei et al.~\cite{wei2022clear} propose CLEAR, an API recommendation tool based on contrastive learning. It employs the RoBERTa model for natural language modeling with a contrastive learning objective, where the model learns to determine whether two given inputs are similar instead of focusing on learning specific characteristics of each classification category. Based on previous observations, language models are predominantly used to model the natural language aspects of API recommendation questions.

\mybox{Answer to RQ4}{red!40}{gray!10}{
\small{
There are three common modeling approaches in API recommendation i.e. statistic model, machine learning model, and deep learning model. The most popular statistic models are the TF-IDF model and the Indexing Model. The most popular machine learning model is the Word2Vec model, and the most popular deep learning models are LSTM, CNN, and BERT models.
}
}
\section{RQ5: What are common evaluations in the API recommendation}
\label{RQ5}
\begin{table}[t!]
\caption{Common Evaluation Metrics }
\label{table_18}
\begin{tabular}{cll}
\toprule
No & Metrics &  Reference      \\
\midrule

1 & Accuracy &\cite{nguyen2017exploring,yan2018learning,xie2020api,liu2018effective,nguyen2016api,sworna2023apiro,kang2021apirecx,chen2021novel,tian2018automatically,chen2021holistic,lee2021mining,rahman2016rack}\\
2 & MRR  &\cite{huang2018api,li2020fusion,zhou2021boosting,liu2018effective,wei2022clear,sworna2023apiro,chen2021holistic,lee2021mining,rahman2016rack,wang2021hybrid,zhang2017recommending}\\
3 & MAP &\cite{huang2018api,li2020fusion,zhou2021boosting,lee2021mining,sworna2023apiro,rahman2016rack,wang2021plot2api,wei2022clear,zhang2017recommending}\\
4 & BLEU &\cite{tian2018automatically,gu2016deep,irsan2023picaso}\\
5 & Hit ratiao &\cite{li2020fusion,zhou2021boosting,zhao2021icon2code,sun2019enabling,zhang2017recommending}\\
6 & Precision &\cite{nguyen2012grapacc,gao2021api,ling2019graph,raghothaman2016swim,wang2021hybrid}\\
7 & Recall &\cite{nguyen2012grapacc,raghothaman2016swim,yuan2019api,liu2019autoencoder,thung2013automatic,liu2019design,wang2021hybrid}\\
8 & NDCG &\cite{wang2021hybrid}\\
9 & F1 &\cite{nguyen2012grapacc,ling2019graph}\\
10 & Success Rate &\cite{heinemann2012identifier,wang2021hybrid,zhao2021icon2code}\\

\bottomrule
\end{tabular}
\end{table}

\subsection{What are the common metrics?}

In addition, we compared the evaluation metrics utilized by API recommendation tools. Since API recommendation is generally a problem involving information retrieval, the evaluation metrics employed by the tools are primarily derived from information retrieval metrics. Table \ref{table_18} displays the evaluation metrics used by these instruments. The next section describes each metric in the table.

\subsubsection{Accuracy, Precision@K, Recall@K or hit ratio@K, and F1@K}
Accuracy, Precision@K, Recall@K, and F1@K are frequently employed metrics for evaluating the performance of a recommendation system. These metrics in API recommendation are typically calculated based on the first K results, where K represents the top K recommendations. Given the answers in a testing set and a model's predictions, the accuracy of a recommendation is the proportion of correctly predicted results relative to the size of the dataset. Precision@K, Recall@K, or hit ratio@K, and F1@K are characterized as follows:

\begin{equation}
Precision@K = \frac{\text{number of top k recommendations that are relevant}}{\text{number of items that are recommended}}
\end{equation}
\\
\begin{equation}
Recall@K = \frac{\text{number of top k recommendations that are relevant}}{\text{number of all relevant items}}
\end{equation}
\\
\begin{equation}
F1@K = 2 \times \frac{Precision \times Recall}{Precision + Recall}
\end{equation}
\\

\subsubsection{Mean Reciprocal Rank(MRR)}

MRR is calculated as the average of the reciprocal rank (RR) across all users. The reciprocal rank, in this context, represents the inverse of the position of the first correctly recommended item. The definition is as follows:
\begin{equation}
MRR = \frac{1}{\left | Q \right |} \sum_{i=1}^{\left | Q \right |}\frac{1}{rank_i}
\end{equation}
Where Q is the length of the test set. MRR is a suitable metric for evaluating API recommendation systems because it is particularly effective when the first correct recommendation holds significant importance. This is typically the case when the first API recommendation result is the most critical API.

\subsubsection{Mean Average Precision(MAP)}
Mean average precision measures the mean of average precision (AveP) of the test result. The formula of AveP and MAP is defined as the following:
\begin{equation}
AveP@K = \frac{1}{m}\sum_{k=1}^{N}(P(k) \text{ if } k^{th} \text{ item was relevant})
\end{equation}
\begin{equation}
MAP = \frac{\sum_{q=1}^{Q}\text{AveP}(q)}{Q}
\end{equation}
Where Q is the length of the test set. The MAP metric primarily assesses the comprehensiveness of recommendation results. MAP is a valuable metric in the API recommendation field as it is frequently employed to gauge the completeness of API recommendation results.

\subsubsection{Bi-Lingual Evaluation Understudy(BLEU)}
BLEU is a widely accepted metric for evaluating the performance of language models in natural language processing. Originally designed to assess the quality of machine-translated text, it has since been extended to evaluate the quality of sequences generated by models. The BLEU score ranges from 0 to 1, indicating the similarity between the translation and the reference translation. In API recommendation, BLEU is employed to measure the similarity between the generated API sequence and the reference sequence, considering both the number of mathematical APIs and the sequence of tokens.

\subsubsection{Normalized Discounted cumulative gain (NDCG)}
Similar to BLEU, NDCG assesses the quality of a ranked list. The Cumulative Gain (CG) is calculated as the sum of all relevant results within the sequence's length. Discounted Cumulative Gain (DCG) enhances CG by applying a logarithmic function to consider the sequence order. To account for varying sequence lengths and to normalize the score within the range of zero to one, Normalized Discounted Cumulative Gain (NDCG) is used. However, NDCG is less commonly utilized in evaluating API recommendation results, particularly in cases where it is insensitive to false positives.

\subsubsection{Success rate}
Success rate measures the rate of at least one success match in the top K result. For example, For a dataset P, the formula of success rate is defined as:
\begin{equation}
\text{success rate}(p) = \frac{\sum_{i}^{N}match_i(p)}{\sum_{i}^{N}\left | G_i \right |} \times 100%
\end{equation}

\mybox{Answer to RQ5}{red!40}{gray!10}{
\small{
Among the eleven evaluation metrics, Accuracy, MRR, and MAP are the most commonly accepted metrics. Precision, hit ratio, and recall are also very popular options for API recommendation evaluations.

}
}
\section{challenges}
\label{challenges}

\subsection{Challenge 1: Lack of Benchmarks}
The absence of a benchmark is the first and most obvious challenge in API recommendation. While a few studies reuse existing datasets for performance comparison, the vast majority of existing studies create their own datasets. The reason for this is that different tasks require various types of information, and each tool serves a unique purpose. Additionally, software development is characterized by versioning, making it challenging to reuse existing datasets. Instead, researchers often opt to collect the most recent version of a data source using the same procedure as before, prioritizing data relevance. Developing a unified benchmark encompassing the diverse data attributes used in the dataset collection process would be formidable.

\subsection{Challenge 2: Dataset Quality Issues}

The quality of datasets is the second challenge in API recommendation research. Specifically, validating the ground truth of datasets is not easily achievable through automated methods alone. In some cases, manual verification is required to filter the test set and confirm the ground truth of the data~\cite{huang2018api,campbell2017nlp2code}. This issue becomes even more prominent in the case of query-to-API style API recommendation.

The training of a query-to-API style API recommendation tool involves using data pairs consisting of programming queries and API responses. Stack Overflow is the primary data source for this type of dataset, as researchers assume that questions on Stack Overflow predominantly pertain to how-to questions related to API recommendation. However, this assumption doesn't always hold true, as there are other types of questions, including those related to bug-fixing and performance comparison. This inconsistency poses a risk that the ground truth of query-to-API API recommendation tools may not always be guaranteed.

Furthermore, many code snippets require context for the API call to function correctly, but the APIs mentioned in the context may not always be relevant to the question. It's often the case that only one or two API methods actually solve the query's problem. Identifying and isolating the relevant context API from the essential API can also be challenging.



\subsection{Challenge 3: Lack of Measurement Variety}
Most existing studies train source code models from scratch for source code modeling. This practice is not efficient compared to the pre-training and fine-tuning paradigm in natural language processing. Existing studies borrow pre-trained deep learning models from the natural language processing field and repurpose them for source code modeling through fine-tuning, even though there are not many pre-trained models specifically for source code. Training large deep-learning models is costly, and CodeBert is one of the few pre-trained models for source code modeling that is actually useful.

\subsection{Challenge 4: Lack of language variety}

Out of the 35 collected articles, 20 studies focus on Java API recommendations. Only a limited number of studies are dedicated to other languages, such as Python, JavaScript, and C. No studies can be found on languages like PHP, Go, or Kotlin. Considering that each programming language has its own strengths, most accepted use cases, and community, there may be a gap in API recommendation results between different languages. Conducting such a study would be worthwhile, but it remains open for exploration.

\subsection{Challenge 5: Lack of Practical Usability Evaluation}

After analyzing all articles, we have observed that most studies approach API recommendation tasks as sequence generation tasks. This approach focuses more on simulating the sequence rather than deeply understanding the actual needs of the user and the API's functionality. An ideal model should recommend APIs based on their semantic and actual features, rather than just simulating an API sequence.
Although there have been several graph-based approaches attempting to construct a graph representation of API knowledge and map APIs to user queries, a comprehensive approach to API knowledge representation is still lacking. This is necessary to recommend APIs based on a deep understanding of their features.




\section{Opportunities}
\label{Opportunities}

\subsection{Opportunity 1: unified dataset.}
To improve the current state of API recommendation, a unified dataset should be created and used as the benchmark. This dataset should include information from various sources used in existing API recommendation research. The dataset should contain as much of the project as possible for code completion-style API recommendation, along with information from code analysis such as abstract syntax trees, control flow graphs, and method call sequences. On the other hand, for query-to-API-style API recommendation, the dataset should only contain questions that request API problem solutions, such as "How-to" questions. A middle representation like LLVM should also be included to enable translation between multiple programming languages.

\subsection{Opportunity 2: COntinous Learning of Pre-training Models}

The use of pre-trained models is an essential technique for general language modeling. Such models are trained using vast amounts of data and parameters, only once, and can then be fine-tuned for multiple downstream tasks. In the field of API recommendation, Feng et al.~\cite{feng2020codebert}introduced CodeBERT, a pre-trained source code model for generating code in different programming languages. CodeBERT has been evaluated for a variety of software engineering tasks and consistently delivers impressive performance. However, due to the emergence of new libraries and software APIs, pre-trained models can become outdated quickly, and training new models frequently can be costly. Therefore, a continuous learning approach is necessary to keep the model up-to-date and aware of the latest libraries and APIs.

\subsection{Opportunity 3: Different Modeling and Training Paradigm}
In recent decades, research in natural language processing has undergone rapid evolution. A relatively new paradigm, known as prompt learning, has emerged in the field of NLP. This paradigm generates predictions by completing a sentence referred to as a "prompt." It effectively transforms classification tasks into masked sentence completion tasks, maximizing the potential of pre-trained masked language models. However, adapting this paradigm to current API recommendation models is not straightforward due to the requirement of a large pre-trained code model, which is scarce in the API recommendation field.
\section{conclusion}
\label{conclusion}
This survey presents a systematic literature review focusing on the API recommendation task. we have collected and analyzed 34 papers in the API recommendation field. Based on the empirical findings of current approaches, we draw some conclusions. We identified several obstacles in the current state of API recommendation research. We found that 1) the API recommendation task lacks a unified benchmark, 2) obtaining a high-quality API recommendation dataset is still expensive, and 3) the topic of the pre-trained large code model is understudied compared to the progress made on the pre-trained language model. We also found that the evaluation of usefulness is difficult to automate and requires manual effort at present. We believe that the future of API recommendation will focus on a unified benchmark, a larger pre-training code model, and new paradigms for API learning, such as few-shot learning and prompt learning.


\bibliographystyle{ACM-Reference-Format}
\bibliography{sample-base}

\appendix

\end{document}